\documentclass[12pt]{article}

\usepackage{scicite}
\usepackage{array}

\usepackage{times}
\usepackage{tabularx}
\usepackage{tablefootnote}
\usepackage{hhline}
\usepackage{graphicx} %
\usepackage{todonotes}

\newenvironment{sciabstract}{%
\begin{quote} \bf}
{\end{quote}}

\newcounter{lastnote}

\title{The cost of reading research. A study of Computer Science publication venues}

\author
{Joseph Paul Cohen, Carla Aravena, Wei Ding\\
\normalsize{Department of Computer Science, University of Massachusetts Boston, Boston, MA, USA}\\
}

\date{}
\begin{document} 
\baselineskip24pt
\maketitle 
\begin{sciabstract}
\end{sciabstract}

\section*{Introduction}

What does the cost of academic publishing look like to the common researcher today? Our goal is to convey the current state of academic publishing, specifically in regards to the field of computer science and provide analysis and data to be used as a basis for future studies. We will focus on author and reader costs as they are the primary points of interaction within the publishing world.
In this work, we restrict our focus to only computer science in order to make the data collection more feasible (the authors are computer scientists) and hope future work can analyze and collect data across all academic fields.

Today, there is an echo of the decade-old concerns of Knuth, Jordan, and Odlyzko regarding publishers that pose an unnecessary cost burden on academic readers. In 2003, Donald Knuth questioned the subscription cost to Journal of Algorithms, which was increasing year after year (adjusted for inflation), causing the price per page to double from 1980 to 2003 \cite{knuth_letter_2003}. Michael Jordan along with 40 other members of the editorial board of the Machine Learning Journal (MLJ) famously resigned in 2001 in protest of papers being locked behind the paywalls of its publisher \cite{jordan_leading_2001}. Similarly, Andrew Odlyzko discussed how technology eliminates the need for middlemen and publishers, claiming that librarians were an unnecessary block between scholars and their audience. Because of this, he argued that publishers and librarians would have to defeat free access preprint electronic distribution in order to stay relevant \cite{odlyzko_tragic_1995}.

In response, publishers frequently propose the idea that such fees, whether placed upon the readers or the authors of the papers, are necessary in order to ensure quality content. Allowing for a proper peer review, proofreading, and causing widespread circulation of the paper are some of the reasons publishers cite for the necessity of fees. We study this in \S \textbf{Author and Reader Costs} and find that there is no correlation between the influence of a venue and the cost to the authors or reader. 

Knuth recognizes that, in the past, the publisher's role in ``keyboarding and proofreading'' was valuable (and expensive), today ``authors have taken over most of that work, and software has also ameliorated the other aspects of a publisher's task'' so that nearly anyone with access to certain software (like \LaTeX) can produce a high-quality and visually appealing paper \cite{knuth_letter_2003}. We explore this in \S \textbf{Author and Reader Costs} and conclude that the more influential venues are those with free paper access. In \S \textbf{Cost and Influence} we rule out sponsorship after finding no correlation between the number of sponsors and influence of the venues.

Today, the issue has found new momentum with voices such as Peter Suber discussing all aspects of the issue in \cite{suber_open_2012} and focusing on changing the publishing policies of university faculty through incentives. Recent analysis by Schmitt \cite{schmitt_academic_2014} discussed the lucrative profits of the publishing companies. Schmitt described how costs of 3- to \$3.5-million annually paid by universities are pushing even well-established university libraries such as Harvard University to state that they ``can no longer afford to pay for all the journal subscriptions.'' There are now various open access models which offer free access with the caveat that they charge authors to publish their papers instead. Solomon \cite{solomon_study_2012} performed a survey of open access publishers and found the cost at between \$8 and \$3,900 to publish articles in all fields. Laakso \cite{laakso_anatomy_2012} has analyzed the adoption of open access to find that the annual share of publications is increasingly open access--specifically biomedicine--due to mandates for research funded by the U.S. National Institute of Health. However, there are many concerns on the legitimacy of these venues as discussed by Butler \cite{butler_investigating_2013} who criticized some of these venues for ``shady publishing practices'' calling many of them ``potential, possible or probable predatory scholarly open-access publishers.''

With these concerns in mind, it is important to study where the field of computer science is in terms of the cost for both readers and authors. We create a list of top computer science venues based on their h5-index, collect author and reader costs as well as other attributes, and attempt to look for patterns.

Many interesting observations were found in this study. First off, in this analysis we find no significant pattern, in either direction, between cost and influence of venues, meaning cost may not imply influence.  

Also, many journals have a high reader or author cost, leading us to believe cost must come from one or the other, however there are many high ranking venues which have neither of these costs and draw their funding from other sources. Perhaps the most interesting observation in this paper is that, in every subfield of computer science, the number of free access venues is not proportional to their h5-index influence. The ratio of free access venues to non-free access venues is always less than the ratio of the total h5-index of free access venues over the total h5-index of non-free access venues.

\section*{Data}

Our goal is to study the top conferences in computer science and paint a picture of the current state of publishing in computer science with regard to reader and author costs. For this task, we used Google Scholar's rankings for Top Venues in Computer Science\footnote{https://scholar.google.com/citations?view\_op=top\_venues}. Google Scholar provides public venue data covering traditional conferences and journals as well as non-traditional open access and free access venues (such as $ar\chi iv$\footnote{http://arxiv.org/}) that are not commonly included in venue rankings. The venues in this data are ranked by $h5$-index which provides the data we need to measure the influence of these venues.

We study the top 20 venues from each subfield related to computer science resulting in 288 venues after removing duplicate entries. The $h5$-index is used to rank each venue as opposed to impact factor because of its tolerance to noise and because it is available in the data source. The $h5$-index is the $h$-index for articles published in the last 5 years. The h-index of a venue is the largest number $h$ such that $h$ articles published in the last five years have at least h citations each. The impact factor is the number of citations to articles published in a venue for a given year. For example, if a venue publishes 50 papers and none of them receive a citation except one which is cited 1000 times then the impact factor would be 20 while the h-index would be 1. However, publishing papers that are not cited can impact this score negatively where the h-index would not be affected. The $h5$-index captures the strength of the authors choosing that venue to disseminate their work which aligns more with our goal in finding the venues with the highest influence.

For each venue, we sent a survey to the editors/organizers asking for information as well as manually garnering information from their associated websites. An overview of the dataset that will be made public is in Table \ref{tab:attribs}. The reader costs are the non-subscription prices which are the prices for purchasing a single article. The author costs are the minimum costs for a non-student to have their paper published. This includes conference registration costs as well as any journal publication fees.

\section*{Author and Reader Costs}
\label{sec:costs}

In this section we study how reader and author costs are distributed in relation to their h5-index. In the first analysis we find there is no significant correlation between h5-index and reader or author cost. Additionally, grouping of conferences around cost and h5-index can be seen linked with publishers. This section then explores the outliers of the data in terms of cost and impact.

\begin{figure}[!H]
\begin{center}
\includegraphics[width=0.99\textwidth]{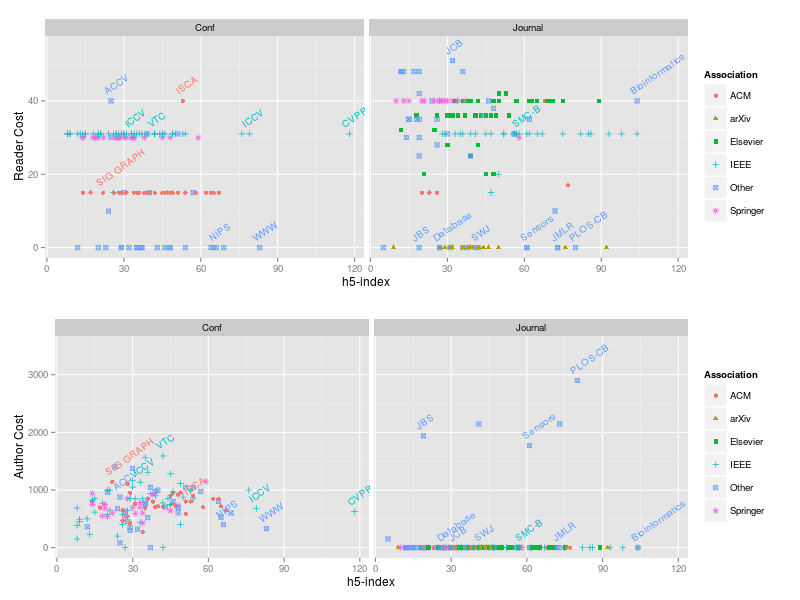}
\caption{The influence of each venue and the cost to the reader and author are compared. The venues are colored based on publisher to show the clustering of their prices and influence. The ``Other'' association contains smaller publishers as well as foundations. Venues are split based on if they are a journal or a conference.}
\label{fig:cost-h5}
\end{center}
\end{figure}

In Figure \ref{fig:cost-h5} cost and h5-index are plotted against each other split between conference and journals. We find specific clusters of venues with very similar reader costs. When each point is colored by association it's apparent that this is the reason. Publishers like ACM and IEEE have a large market share and charge a flat rate for each paper. The IEEE charges \$31 per article for almost all conference and journal articles and ACM similarly charges \$15 per article. For author cost, which is highly variable based on the location of the conference, it appears there is no pattern which resembles the grouping that appears in the reader costs.

It is interesting to note the juxtaposition of some venues which charge differently yet have similar rank. For example, while Bioinformatics and PLOS-CB (Public Library Of Science Computational Biology) have similar h5-index values, they are on the opposite end of the reader cost and author cost spectrum. This can also be seen again between the journals Sensors and SMC-B (IEEE Transactions on Systems, Management, and Cybernetics, Part B). This would appear to be a fundamental pattern of journal financing implying that cost must come from either readers or authors. However, there are many outlier journals that break this pattern and charge nothing for author and reader including JMLR (Journal of Machine Learning Research), SWJ (Semantic Web Journal), Databases, and all arXiv venues.

Why do these outliers exist? JMLR was famously created by editors who resigned from Machine Learning Journal (MLJ) over how the journal publisher was "restricting the communication channel between authors and readers" \cite{jordan_leading_2001}.  Stuart Shieber, a computer science professor at Harvard explains how JMLR can afford this in a Blog Post \cite{shieber_efficient_2012}. Given the prominence of the editors it was not difficult to solicit publications, typesetting is done by authors themselves using \LaTeX, reviewing is a volunteer effort as always, and website hosting is taken care of by MIT. Shieber states the largest cost is in hiring a tax accountant.

Another outlier is Cornell's arXiv, a well known preprint service funded by ``Cornell University Library, the Simons Foundation and by the member institutions.'', which charges nothing to the author and reader. ArXiv itself does not peer review papers but some conferences use this service to host their papers. Conferences such as the International Conference on Learning Representations (ICLR) request that users submit their papers to arXiv before submitting their paper for review which results in free access to papers after they are accepted.

SWJ, JMLR, and arXiv have a common feature: the author retains the copyright for the work and the venue only has a license to distribute it. However, this is not the case for every free-to-read venue. It is also interesting to note that some of these venues labeled free access have restrictive copyright agreements. The proceedings for WWW are available for free download on the conference website but are also sold for \$15 on the ACM Digital Library. The copyright terms\footnote{http://wwwconference.org/proceedings/www2014/starthere.htm} state that the articles are free for personal and classroom use only, otherwise a fee must be paid for reproduction. This restrictive license potentially allows the copyright holder to cease free distribution and rely exclusively on the ACM Digital Library.

\section*{Cost and Venue Influence}
\label{sec:influence}

In this section we study relationships between the cost of an article to a reader and author and the influence of that venue. We call a venue with a zero cost to the reader as free access and a non-zero cost as non-free access, the same with free publish and non-free publish works. First, we look at the associations of conferences and journals and the cost to read them. We then analyze the distribution of conferences that offer free access to papers within each field. We find that conferences with free access to read have higher influence in every field in proportion to the number of venues in those fields. We also find that there are three fields that do not have any top conferences that are free access.

\begin{figure}[!H]
\begin{center}
\includegraphics[width=1\textwidth]{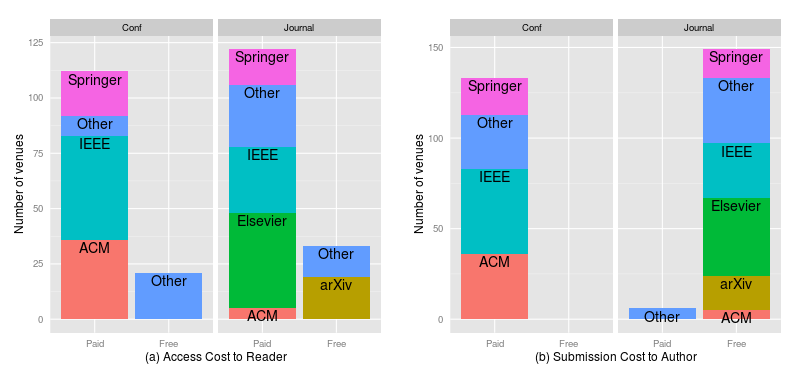}
\includegraphics[width=0.90\textwidth]{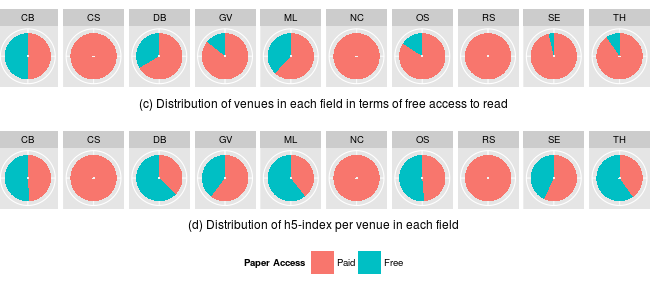}
\caption{(a) and (b) show an overview of author and reader costs in conference and journals. Large publishing groups are labeled by color.  (c) and (d) show the data for conferences broken down into fields. Each field is divided into those which provide papers for free and those that require payment.}
\label{fig:conf-dist-pie}
\end{center}
\end{figure}

We study the distribution of venue costs between conferences and journals in Figure \ref{fig:conf-dist-pie}a and \ref{fig:conf-dist-pie}b. The distribution is very disproportional and none of the major publishers offer anything that is free to read in either venue type. No conferences in our dataset offer free submission while almost all journals do. Computer Science does not have many open access journals with top ranking.

To delve deeper into this analysis we break down venues into sub-fields as shown in Table \ref{tab:fields} based on the fields used by Osmar R. Zaiane in his conference ranking site\footnote{http://webdocs.cs.ualberta.ca/\textasciitilde zaiane/htmldocs/ConfRanking.html}. There are many similar lists of subfields in computer science but this list seemed the most complete. However, many of the venues in our dataset needed to be classified manually because they were absent in this list due to the larger number of venues that were included in Google's list. Also, some fields were too sparse to be worth plotting and were merged into other categories.

We discovered a very interesting pattern in the influence distribution of conferences. When there are free access conferences in a field, the proportion of free access to paid access conference was always lower than the proportion of their representation in the overall cited papers (based on their h5-index). To illustrate this we plotted the proportion of free access to paid access conferences in Figure \ref{fig:conf-dist-pie}c. 100\% of the circle represents the total number of venues in that field and the colored sections represent each type of access. Every field has less free access conferences than non-free access. Figure \ref{fig:conf-dist-pie}d shows the distribution of venues based on their h5-index. 100\% of the circle represents the sum of all h5-index values for that field and the colored sections represent the access type of the venue that provided that h5-index.

This pattern is most notable in Databases, Theory, Graphics, Security, and Operating Systems where the difference is more than double. General Computer Science, Networking, and Remote Sensing have no free access conferences and a low number of conferences in general. Some of the fields with the highest proportion of free access papers are Computational Biology, Databases, Machine Learning, and Theory. We can speculate that the higher influence is due to these conferences receiving more exposure because their papers are more easily accessible.

\begin{figure}[!h]
\begin{center}
\includegraphics[width=1\textwidth]{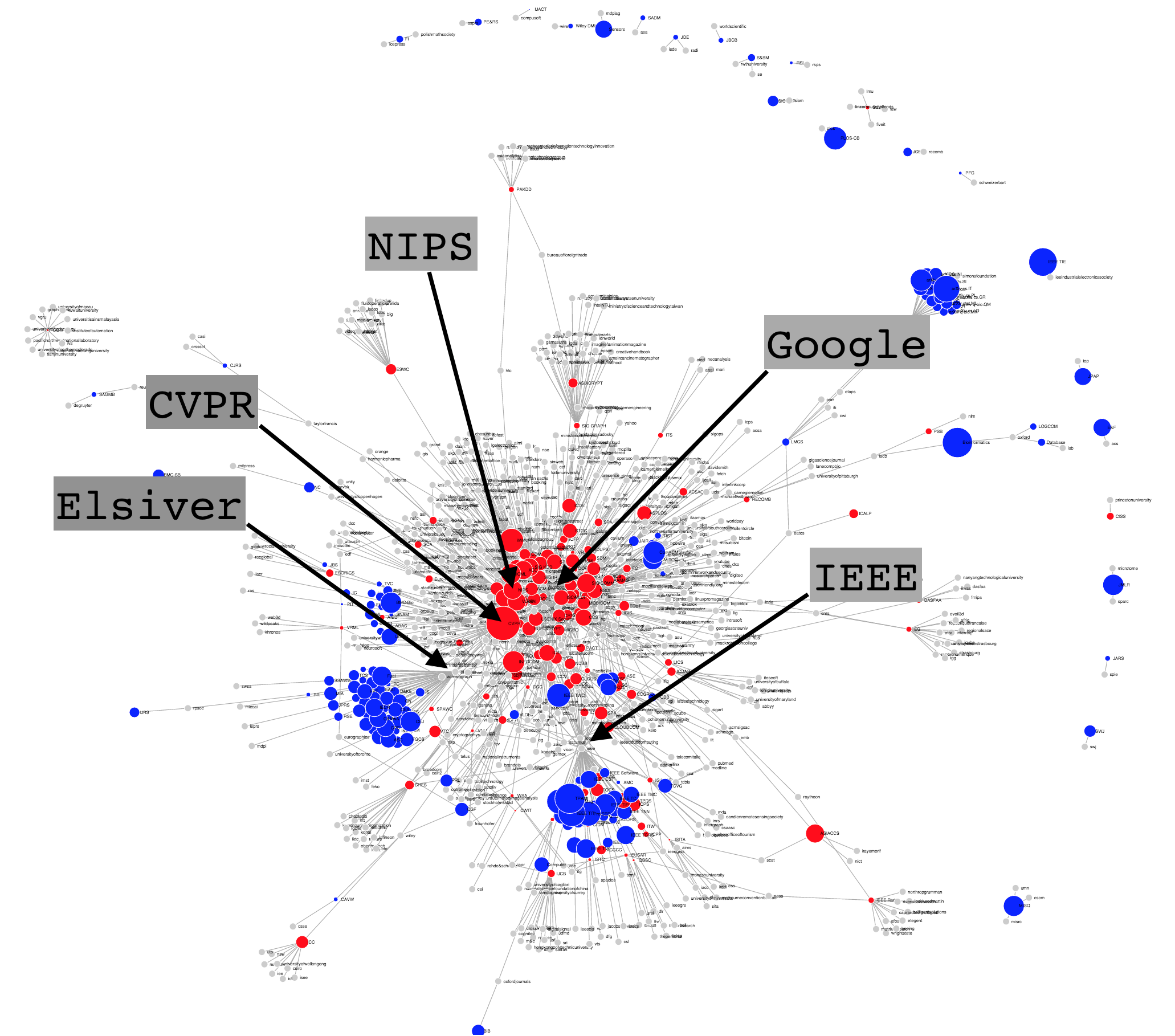}

\includegraphics[width=0.48\textwidth]{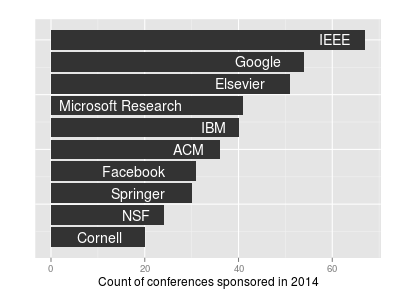} %
\includegraphics[width=0.48\textwidth]{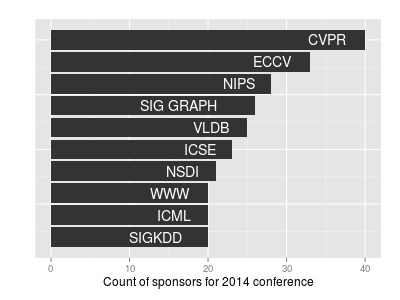}
\caption{The relationships between sponsors and venues. The nodes are colored as follows: (Red, Conference), (Blue, Journal), and (Grey, Sponsors). The size of the nodes represent their h5-index score. Edges represent a sponsor relationship. The sponsors give money to the conferences and journals. A live representation can be seen here: http://www.cs.umb.edu/\textasciitilde joecohen/csvenue/sponsorgraph}
\label{fig:sponsorgraph}
\end{center}
\end{figure}

Next we analyze the relationship between venues and sponsors. A graph is constructed linking sponsors to venues by creating nodes for sponsors and nodes for venues. An edge is added for each sponsor relationship. An overview of this graph is shown in Figure \ref{fig:sponsorgraph}. Plots of the top 10 highest degree nodes are shown below the graph depicting IEEE and Google as the groups that sponsor the most venues analyzed. There seems to be no relationship between whether or not the venue has free access to their papers. CVPR, ECCV, and SIGGRAPH do not offer free access while NIPS and VLDB do offer free access.

\section*{Conclusion}

With the possibility of fake companies seeking to make money off of authors for publishing their papers, increasing access prices possibly limiting the number of potential readers, the fact that much of the editing of articles can now be easily done by the authors' themselves, along with our data suggesting that there is not only no significant correlation between our measure of a paper's influence and its reader or author cost, the necessity of such high fees by publishing companies comes into question. While more research could be done on this topic across different years, we see this research as a starting point for more data to be acquired on computer science publications and the impact of their ever-increasing fees for readers and authors. We hope this data can be used in the future to observe the progression of cost in academic publishing and to ensure a future without lost research due to monetary restrictions.

\begin{table}[ht]
\centering
\begin{tabular}{c|>{\centering\arraybackslash} m{10cm}}
Attribute & Description \\
\hline
\hline
Field & In Table \ref{tab:fields} \\\hline
Conf Journal & If the venue is a Conference or Journal\\\hline
Abbrev & Abbreviation of the conference \\\hline
Conference & Full conference name \\\hline
Data Year & Year that data was collected \\\hline
Conf Chairs/ Editor & Names and emails of conference chairs and editors\\\hline
h5-index & Over the past 5 years, the mean of how many citations resulted from papers published in this venue \\\hline
h5-median  & Over the past 5 years, the median of how many citations resulted from papers published in this venue \\\hline
\# Submitted Papers & Given or calculated from provided data\\\hline
\# Accepted Papers & Given or calculated from provided data \\\hline
Acceptance Ratio &  Given or calculated from provided data\\\hline
Travel Grants & If travel grants are given \\\hline
Travel Grant Funding & \\\hline
Author Cost (\$USD) & Minimum cost for a paper to be published. Includes conference registration or publication fee \\\hline
OA Prices (\$USD) & Price for publishing a paper open access in a typically subscription only venue\\\hline
Reader Cost (\$USD) & Individual cost of a paper purchased from a paywall\\\hline
Association & If the venue is part of ACM, IEEE, SIAM, etc \ldots \\\hline
Impact Factor & Total papers cited / total papers published\\\hline
Plans for Free Access? (Y/N) & If the venue has \\\hline
\# Attendance & How many people attended the last conference\\\hline
Conference Cost & How much the conference cost to organize\\\hline
Sponsors & The names of previous years public sponsors\\\hline
\end{tabular}
\caption{Attributes attempted to collect for every venue}
\label{tab:attribs}
\end{table}

\begin{table}[ht]
\centering
\begin{tabular}{c|c}
Field & Description \\ 
\hline 
\hline
 TH & Theory\\ \hline
 SE & Security and Privacy\\ \hline
 RS & Remote Sensing\\ \hline 
 OS & Operating Systems / Simulations\\ \hline 
 NC & Networks, Communications\\ \hline
 ML & Machine Learning\\ \hline
 GV & Graphics, Vision and HCI\\ \hline 
 DB & Databases \\ \hline 
 CS & General Engineering and Computer Science\\ \hline
 CB & Computational Biology\\ \hline

\end{tabular}
\caption{Fields used to label each venue. Extended from Osmar R. Zaiane’s Conference Rankings}
\label{tab:fields}
\end{table}

\bibliography{csvenue}

\bibliographystyle{Science}

\clearpage

\end{document}